# Duty to Delete on Non-Volatile Memory

N.Y. Ahn [1*], D.H. Lee [2**]

[1] Graduate School for Information Management, Korea University, 145, Anam-ro, Seongbuk-gu, Seoul, Korea
[2] Professor at CIST and Graduate School for Information Security, Korea University
[*] humble@korea.ac.kr  [**] donghlee@korea.ac.kr

**Abstract:** We firstly suggest new cache policy applying the duty to delete invalid cache data on Non-volatile Memory (NVM). This cache policy includes generating random data and overwriting the random data into invalid cache data. Proposed cache policy is more economical and effective regarding perfect deletion of data. It is ensure that the invalid cache data in NVM is secure against malicious hackers.

## 1. Introduction

On, May 13, 2014, the Court of Justice of the European Union (CJEU) announced the historical decision in personal information protection, which is "the right to be forgotten" in the context of data processing on internet search engines [1], [2]. CJEU decided that the internet service providers (ISPs) of the search engines would be responsible for the processing of personal information in web pages by third parties [3]. Recently Sachiko Kanamori, Kanako Kawaguchi, and Hidema Tanaka introduced the scheme for the right to be forgotten using secret sharing and digital watermarking in social networking services (SNSs) [4]. And Hiroki Yamazawa, Kazuki Maeda, Tomoko Ogura Iwasaki and Ken Takeuchi at Chuo University proposed privacy-protection solid state storage (PP-SSS) system for internet data's "the right to be forgotten", in which data lifetime is specified without file system overhead [5]. The PP-SSS controls data lifetime using precision error correction code (ECC) and crush techniques. Naturally, the right to be forgotten in system memories should be considered. Especially, if system memories are configured to include a non-volatile memory [6], the internet providers ISPs should design the system memories to meet "the duty to delete" in the non-volatile memory, overwhelming "the right to be forgotten".

## 2. Related Works

Recently, hybrid main memory systems include a central processing unit (CPU), a volatile memory such as a dynamic random access memory (DRAM) and a non-volatile memory (NVM) such as a NAND flash memory, a phase change memory (PCM), a spin transfer torque random access memory (STT-RAM), a ferro-magnetic RAM (FeRAM), etc referring to Fig. 1 [7], [8], [9], [10]. CPU may access caches of DRAM in processes. If a cache is not used by CPU for a predetermined time, the dirty cache is flushed to NVM according to a cache policy [11], [12], [13]. That is, DRAM flushes the dirty cache to NVM. Then the dirty cache still remains in NVM. After flushing, the dirty cache may be updated by CPU. Then, how the corresponding cache in NVM is managed? Firstly, CPU sends invalid information to NVM. Then in a cache table in NVM, a valid/invalid bit of this cache is changed according to the invalid information such as an invalid request from CPU. If the invalid information is personal information, this invalid request can be de-identification request. The de-identification request may be read command or write command with regard to pseudonymization, aggregation, data reduction, data suppression, data masking. Etc.

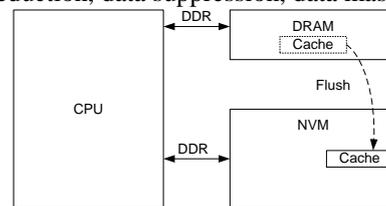

*FIG.1* Hybrid Main Memory

Fig 2. shows an exemplary flush process. Referring to Fig.2, in a flush process, Cache 5 and Cache 6 are moved to NVM. Then, if Cache 5 and Cache 6 are updated by CPU, the updated Cache 5' and Cache 6' are in DRAM. And Cache 5 and Cache 6 in NVM should be invalid. Accordingly, CPU transfers an invalidation request about Cache 5 and Cache 6 to NVM. NVM may manage Cache 5 and Cache 6 as invalid caches in response to the invalidation request. For example, in a cache table, valid/invalid bit corresponding to Cache 5 and Cache 6 are converted.

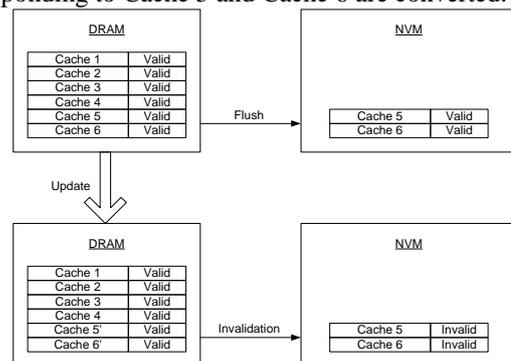

*FIG.2* Flush Operation

For example, in a cache table, valid/invalid bit corresponding to Cache 5 and Cache 6 are converted. That is, the valid bit is converted to the invalid bit according to the invalidation request. Then, it is our important interest how Cache 5 and Cache 6 in NVM are managed. Immediately, Cache 5 and Cache 6 can be physically deleted. But, in general schemes, the above work cannot be generated. This is because the physical deletion is too expensive in regard with time, power consumption, etc. In fact, Juniang Shu etc. studied data a remanence experiment on mobile devices: data cleaning, application uninstallation, factory reset [14].

At least 40% data remanence rate of the target deleted files still remains on mobile devices by 9 weeks. If the target deleted files (that is Cache 5 and Cache 6) are personal information, this situation is serious. We have to strongly apply the duty to delete cache data on NVM, referring to FIG. 3.

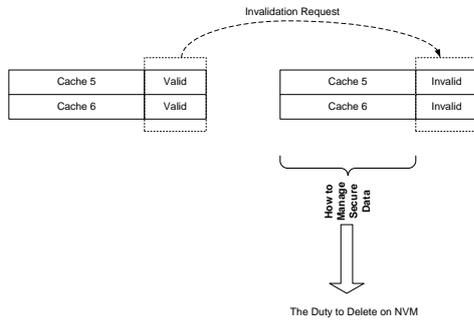

*FIG.3 The Duty to Delete on NVM*

### 3. DDN Process

DDN process is to apply the duty to delete invalid cache data in NVM. We suggest that cache data be overwritten to random cache data in response to the invalidation request in DDN process, referring to FIG. 4. Herein the random cache data are generated in NVM. The generation schemes of random cache data change according to types of NVM. For example, if NVM is an over-writable memory, such as PRAM, MRAM, ReRAM, 3DXpoint Memory, etc., the random cache data may be generated by a random number. Then NVM overwrites the generated random cache data into the corresponding cache data. On the other hand, if NVM is not over-writable memory, such as a NAND flash memory, the random cache data may be only generated in limited environment.

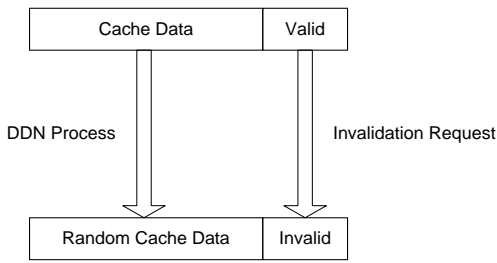

*FIG.4 DDN Process*

In proposed DDN Process, we utilize random data to overwrite invalid cache data, referring to FIG. 5. As a result, the invalid cache data having secure data is changed into random cache data without physical deletion in NVM. We expect that invalid cache data will be cancelled economically from NVM.

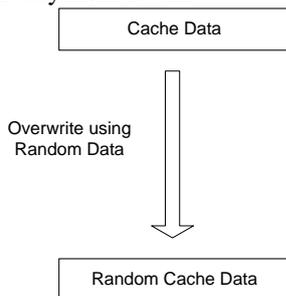

*FIG.5 DDN Process using Overwrite*

### 4. Proposed DDN Process Flow

In general, storage devices include at least one nonvolatile memory NVM and a memory controller NVM CNTL to control NVM [15]. Instead, our proposed NVM CNTL can proceed with DDN process to satisfy the duty to delete on NVM. If the cache is personal information, the DDN process can achieve the privacy protection on NVM. Especially, NVM CNTL includes proposed DDN process unit. The DDN process unit generates random data in response to the received invalidation request from CPU. Herein the invalidation request incudes ADDR corresponding to cache data to be deleted. Then DDN process unit overwrites the random data to a physical space corresponding to ADDR in the nonvolatile memory unit, referring to FIG. 6.

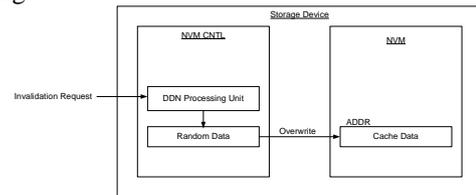

*FIG.6 Proposed NVM CNTL*

Proposed overwriting is largely divided into two points. First overwriting is related to the overwritable memory. In normal random data generation scheme, DDN process unit generates random data by using any random number. Herein the random number may be generated by the random number generator. Then DDN process unit overwrites the random data in the physical space corresponding to the cache data to be deleted from NVM. Second overwriting is related to the non-over writable memory. Also, in limited random data generation scheme, firstly, DDN process unit reads cache data from a nonvolatile memory unit, such as a NAND flash memory. Then DDN process unit generates available random data using the read cache data.

#### 4.1. Cost Comparison

In NAND flash memory, DDN Process includes one reading the original cache data, generating limited random data, and overwriting the limited random data. In terms of time, DDN process takes one reading time, random data generation time, and one write time corresponding to overwriting. Generally, the read/program time is much less than the erase time. For example, for 3D NAND flash memory, page read time is 49 μs, page write time is 0.6 ms, and block erase time 4 ms [16]. For 4 Kbyte, random number generation time is lower than 100 μs [17]. And the erase cost needs to consider the overhead time of garbage collection GC until the erase operation for physical deletion begins. Therefore overwrite cost of DDN Process is less than the erase cost. The proposed DDN Process compared with the erase operation is very economical and effective.

**Table 1** Time Comparison

|  | RD | WR | GEN. time | Erase | GC Overhead |
|---|---|---|---|---|---|
| DDN Process | 49 μs | 0.6 ms | < 100 μs | None | None |
| Prior |  |  |  | 4ms | - |

### 4.2. Available Random Data Generation

If NVM is a NAND flash memory, a number of available states are determined according to an original state corresponding to the original cache data bits. For example, the original state is S5, the number of available states is three (S6, S7, and S8). Random data bits generated by DDN processing unit corresponds one of S6, S7 and S8. Besides, if the original state is the most state S8, the original data bits are maintained. FIG. 7 shows 3-bit random data bits generation in NAND flash memory. If the original state S5 stores the original cache data bits '100', the available random data bits may be one of '101', '110' and '111'.

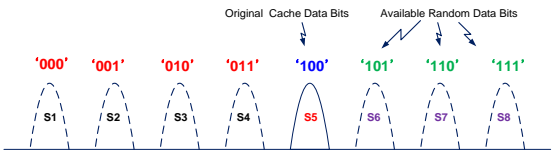

*FIG.7 3-bit Random Data Generation*

### 4.3. Partial Overwrite in a Page

For NAND flash memory, a page includes a plurality of cache data. If at least one of the plurality of cache data is only invalid, DDN Process only overwrites the invalid cache data among the plurality of cache data by a partial program operation. This is called a partial overwrite in DDN Process.

### 4.4. Valid Cache Data applied DDN Process

Although cache data in NVM are valid, the cache data have to be eliminated from NVM after the lapse of predetermined time for security enhancement. If NVM operates in a secure mode, the valid cache data may be overwritten by the above DDN process after the lapse of predetermined time.

### 4.5. DDN process using non-random data

The proposed DDN process does not need to generate random data. Because non-random data may be predetermined for overwriting invalid cache data. It is called to be 'DDN process only data'. For example, DDN process only data may be '1111…11' (size: cache data) corresponding to the most state (eg. 'S8' in FIG.7). This DDN process may omit generating the random data. Referring to FIG. 8, there are many types of the non-random data.

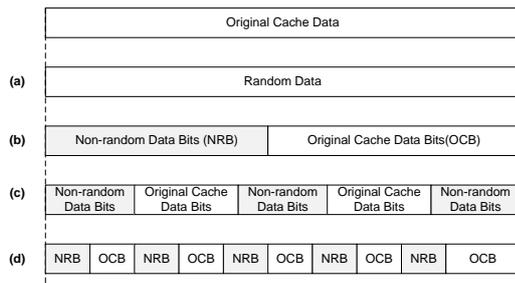

*FIG. 8 Non-random Data Types*

### 5. Conclusion

We introduced new cache policy, that is DDN process applying the duty to delete invalid cache data in NVM. DDN process includes generating random data and overwriting the random data into invalid cache data. Proposed cache policy is more economical and effective regarding perfect deletion of data. Accordingly, DDN Process can protect sensitive cache data such as personal information in NVM against hacking attacks. In future, we need to study new cache architectures to meet "the right to delete" in main memory systems.